\documentclass{article}
\usepackage{ijcai19}

\usepackage{times}
\usepackage{soul}
\usepackage{url}
\usepackage[hidelinks]{hyperref}
\usepackage[utf8]{inputenc}
\usepackage[small]{caption}
\usepackage{graphicx}
\usepackage{amsmath}
\usepackage{booktabs}
\usepackage{algorithm}
\usepackage{algorithmic}

\usepackage{multirow}
\usepackage{subcaption}

\title{BotGraph: Web Bot Detection Based on Sitemap}

\author{
Yang Luo$^1$
\and
Guozhen She$^{1,2}$\and
Peng Cheng$^1$\And
Yongqiang Xiong$^1$
\affiliations
$^1$Microsoft Research\\
$^2$Fudan University\\
\emails
\{yangluo, v-gushe, pengc, yqx\}@microsoft.com
}

\begin{document}

\maketitle

\begin{abstract}
	

The web bots have been blamed for consuming large amount of Internet traffic and undermining the interest of the scraped sites for years. Traditional bot detection studies focus mainly on signature-based solution, but advanced bots usually forge their identities to bypass such detection. With increasing cloud migration, cloud providers provide new opportunities for an effective bot detection based on big data to solve this issue. In this paper, we present a behavior-based bot detection scheme called BotGraph that combines sitemap and convolutional neural network (CNN) to detect inner behavior of bots. Experimental results show that BotGraph achieves $\sim$95\% recall and precision on 35-day production data traces from different customers including the Bing search engine and several sites.


\end{abstract}

\section{Introduction}

According to Incapsula's report \cite{incapsula2016report}, about 51.8\% of Internet traffic in 2016 are performed by automatic bots instead of human. These bots include search engines, price scrappers, Email harvesters and even Trojan which could launch DDoS attacks. These bots not only cause the leakage of business data, but also consume significant bandwidth and server overload. For the past few years, as more businesses migrate their websites to clouds, it becomes the responsibility of the cloud provider to offer an effective bot mitigation solution for its customers.

Based on whether authentication is required, most commonly seen bots can be partitioned into two broad categories: social bots, which target for social networks and web bots, which target for general websites. Compared to social bots, detecting web bots is more challenging because of two reasons. First, it is difficult to identity a website user (or bot) as there is no concept like the user account (or hard to retrieve it as a cloud provider) in the web traffic. Leveraging the client IP address seems to be a feasible method for user identification, Nevertheless, it can be faked easily via proxy. Second, social bot detection can be customized and tuned for a specific social media. However, for web bots, especially as a cloud provider, there would be millions of websites hosted in the cloud. Each site provides distinct services to its customers. Recognizing the bots among all the web traffic for all sites requires a universal scheme that works for hetereogenous scenarios. In this paper, we focus on the detection of web bots. We use the term ``bot'' to refer to web bots in following sections.



There are several traditional ways to perform bot detection, e.g., UserAgent blacklist, IP rate limiting, device fingerprint recognition, etc. However, maintaining IP or UserAgent blacklist requires huge effort to maintain the blacklist database. Moreover, a bot can easily use a proxy IP address or modify its UserAgent to a normal browser. Detecting device fingerprint such as mouse movement and JavaScript engine validation is often a better way, but it usually relies on client-side JavaScript code, which is an invasive technique. Unfortunately, advanced bots can still bypass such detection by utilizing real browser environments like headless Chrome. In summary, all these methods rely on bot's identities or feature codes, which can be easily bypassed by advanced bots via faking their identities. Detecting bots via their behaviors instead of their identities would be a better way.




%

In this paper, we categorize the features of web traffic into two types: identity features and behavior features. Then we introduce a behavior-based bot detection approach called BotGraph. BotGraph is performed in three steps. First, we define the concept of sitemap and propose three ways to build the sitemap for a site. Second, each user session is mapped to a subgraph of the sitemap. The subgraph contains information about which URL patterns the client has visited and the corresponding access frequencies. Third, a 2-dimensional image is generated from the above subgraph. Thus the task of bot detection has become an image classification problem. We use the state-of-the-art techniques like CNN to classify the images into two categories: bot or non-bot. We evaluate BotGraph on various datasets including Bing search engine and several sites from different industries. The result shows that our method can achieve $\sim$95\% precision and recall on most of the datasets.


The remainder of this paper is organized as follows. Section \ref{sec:related_work} elaborates on the related work. Section \ref{sec:approach} presents our behavior-based bot detection scheme called BotGraph. Section \ref{sec:evaluation} brings the experimental results. Section \ref{sec:discussion} discusses about our drawbacks. Section \ref{sec:conclusion} concludes the paper.

\begin{figure*}[!t]
	\centering
	\includegraphics[width=6.0in]{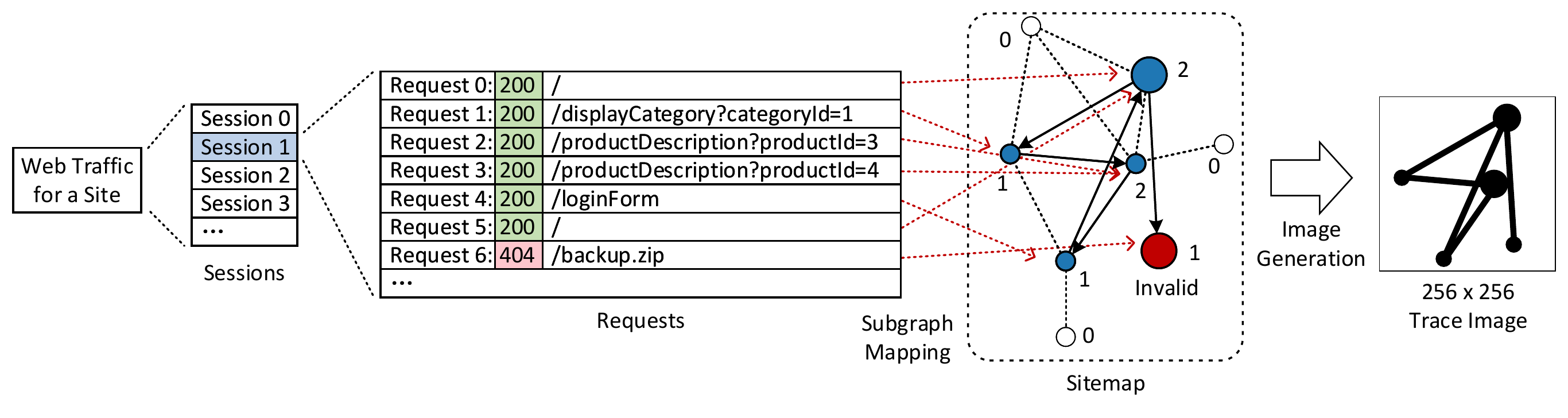}
	\caption{Architecture of BotGraph.}
	\label{fig_approach}
\end{figure*}

\section{Related Work}
\label{sec:related_work}


Bot detection is highly demanded for both website owners and service providers. Google Analytics \cite{googleanalytics} provides built-in filters to help their customers filter bot requests based on geolocation information of the client IP address. Distil, Akamai and ShieldSquare are selected as three leaders in the bot detection area, based on the whitepaper published by Forrester \cite{forrester}. Akamai \cite{akamaiwafintro} uses pre-defined bot signature database as well as a legitimate service whitelist. It also allows its users to customize bot detection rule. ShieldSquare \cite{ShieldSquare} provides both identity-based and behavior-based bot detection. The identity-based method utilizes client-side JavaScript to collect parameters like browser fingerprints. The behavior-based analysis is based on characteristic in terms of number of pages visited per session, time spent on each page, frequency of repeat visits, and so on. This is similar to our solution, but we describe user behaviors in a graph based on sitemap instead, which contains more unique features for a client. Distil \cite{distilintro} provides several bot detection methods such as known violator blacklist, biometric data validation like finger swiping and mouse movement. HTTP's UserAgent is used to recognize the category of the bots. They also provide a machine-learning based method, which needs to be trained for one week before being ready to use. However, UserAgent-based bot detection is not feasible for advanced bots, as they can easily hide themselves by modifying its value.


For feature extraction, some network intrusion datasets \cite{kddcup99intro,csicintro,iscxintro} are provided as the ground truth, some methods were brought out based on which, A review \cite{review} introduced a method which encode the numerical features into holistic metrics like total requests, standard deviative time and the percentage of POST requests, a topic-based model latent dirichlet allocation (LDA) was introduced \cite{ldaevaluation} by Athanasios to encode the semantic information like words in the URLs and postfix representing the type of target resource(eg. html, pdf, asp) into digit feature vectors by introducing the statistic-based concepts like topic variance and topic similarity. Besides that, auto-encoder model \cite{deepautoencoder} is proposed to facilitate the feature extraction procedure. DeepDefense \cite{deepdefense} introduced an algorithm which splited the traffic logs into several segments with the same shape, then encoded the request information in each line of segment to numerical matrix, which would then be fed to recurrent neural network (RNN) based model. Similar to BotGraph, this model can capture the context information among requests. However, the inference efficiency of RNN model highly relies on the length of segments, while BotGraph can perform much more stable.

For detection methods, an improved support vector machine (SVM) algorithm is \cite{oneclasssvm} proposed as the general approach,  More recently, then a boost method called XGBoost \cite{xgboost} is applied to improve both the efficiency and accuracy of detection. Inspired by the trend towards deep learning, the energy-based deep learning model \cite{energymodel} showed up, and GAN-based model \cite{gan} also dabbled in the scope of anomaly detection.

Besides the CNN model used in this work, we also investigated using graph convolutional network (GCN) \cite{kipf2016semi} to train the client's behavior in the web traffic. GCN is different with CNN by accepting structured graphs instead of images as input. However, it is not suitable for our scenario as it can only perform node classification inside one graph instead of classifying multiple graphs. 

Overall, most previous researches rely on identity features such as client IP address or UserAgent for bot detection, which can be easily defeated by advanced bots. A behavior-based bot detection method should be proposed to differentiate advanced bots from normal users.

\section{Our Approach}
\label{sec:approach}

\subsection{Overview}

The architecture of BotGraph is shown in Figure \ref{fig_approach}. There are three steps: first, we need to build the sitemap for the site. We provide three ways to do this. Second, we map the requests in a session to a subgraph of the sitemap. Third, we generate the 2-dimensional trace image from the subgraph, which transforms the bot detection task into an image classification problem. Finally, we use CNN-based model to classify those generated images into two categories: bot and non-bot. We will provide the details of BotGraph as follows. 

\begin{figure}[!t]
	\centering
	\includegraphics[width=2.5in]{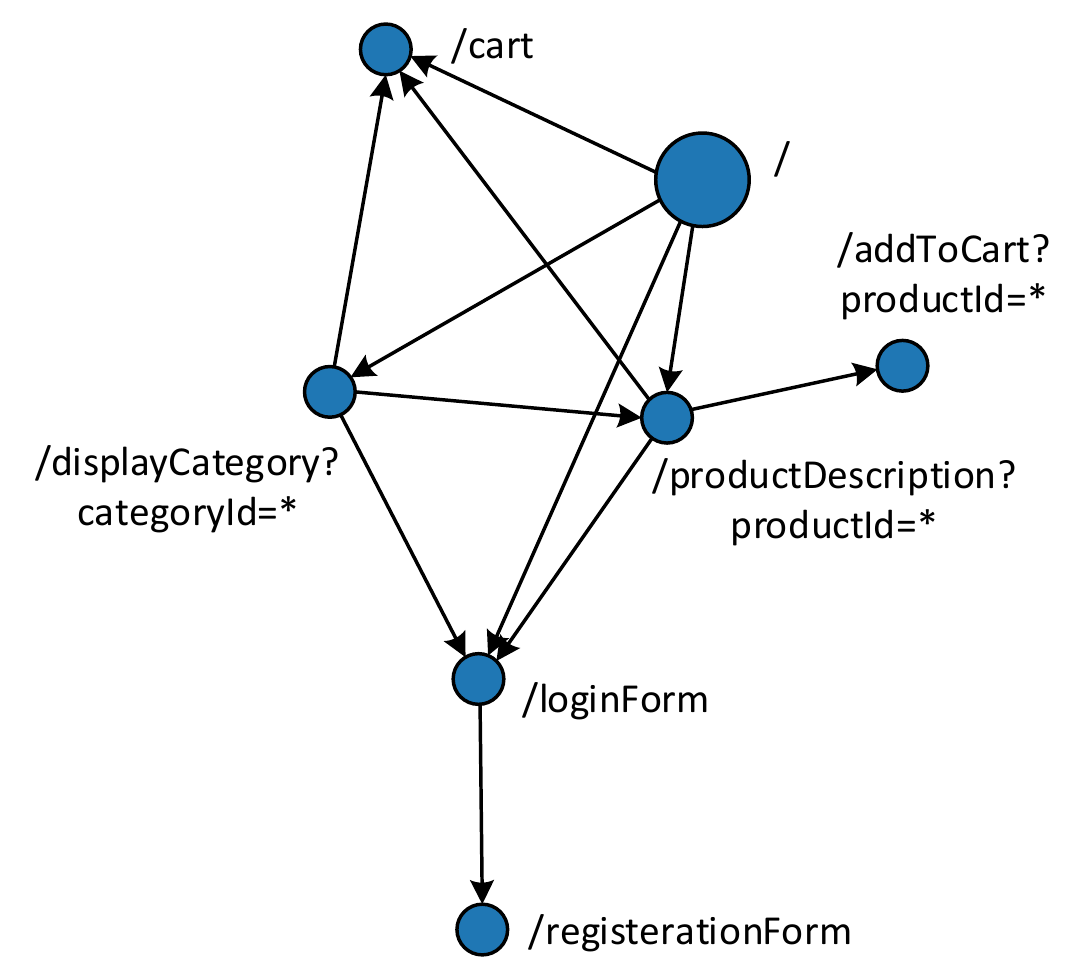}
	\caption{Sitemap of a simple online shopping site.}
	\label{fig_sitemap_example}
\end{figure}

\subsection{Basic Concepts}

\begin{table}
	\centering
	\begin{tabular}{ll}
		\toprule
		Field & Description \\
		\midrule
		Timestamp & Request time, e.g., 2019-1-12 04:00:07 \\
		HttpMethod & HTTP request method, e.g., \texttt{GET}, \texttt{POST}. \\
		RequestUri & The path in URL, e.g., \textit{/books/desc?id=1} \\
		Status & HTTP status code, e.g., \texttt{200}, \texttt{404}.  \\
		Host & ``Host'' field in request header. \\
		UserAgent & ``User-Agent'' field in request header. \\
		ClientIp & Client's IP address. \\
		\bottomrule
	\end{tabular}
	\caption{Request fields.}
	\label{table_request_fields}
\end{table}

\paragraph{Request.} In this paper, we use the term ``request'' to represent the HTTP request from the client to the server (aka the site) together with the corresponding response. A request usually contains many fields, a portion of which is shown in Table \ref{table_request_fields}.

\paragraph{Session.} Bots usually scrape the pages with a large number of requests. To describe such a behavior, a necessary preprocessing step is partitioning the requests into sessions. A session identifies a unique client (normal browser or bot) that performs the accesses.



\subsection{Features}

For the bot detection task, there are two types of fields in a request: identity fields and behavior fields.

\paragraph{Identity fields.} Identity fields are used to identify the client or server. Specifically, \textit{UserAgent}, \textit{ClientIp} are identity fields for the client. \textit{Host} is the identity field for the server.

\paragraph{Behavior fields.} Behavior fields are used to describe the access behavior of the client, including fields like \textit{RequestUri} and \textit{Status}.

Currently, identity fields are widely used in traditional bot mitigation schemes such as IP rate limiting and UserAgent blacklisting. However, if the bots fake their identity through using IP proxy pool, or tamper its \textit{UserAgent}, those methods would fail. Therefore, a feasible way would be detecting the bots via their behavior instead of their identity. The behavior fields used in BotGraph are as follows:






\paragraph{RequestUri, Status.} These two fields play the central role in describing a bot's behavior. We map \textit{RequestUri} of each request into a sitemap node. \textit{Status} is used to determine whether it is a valid mapping. The details would be discussed in Section \ref{sec:subgraph_mapping}.

%



As BotGraph detects bots on a per-session basis, as input, we assume the requests for a site are grouped into sessions and sorted by \textit{Timestamp}. Next we will introduce how we perform bot detection based on sitemap and CNN.

\subsection{Sitemap Retrieval}
\label{sec:approach_sitemap}

Google first introduced the Sitemaps protocol to describe a site's content \cite{google2005sitemap}. A sitemap is a XML file that lists all the URLs for a site. It allows search engines to crawl the site more efficiently. In this paper, we extend the list-formatted sitemap into a graph. The sitemap for a site is defined as: $G = (V, E)$, in which:

\begin{itemize}
	\item $G$: a directed graph.
	\item $V$: set of nodes. Each node represents a URL pattern incorporating multiple URLS. For example, Both \textit{/page?id=1} and \textit{/page?id=2} belong to the same pattern: \textit{/page?id=*}.
	\item $E$: set of directed edges from one node to another. Take two nodes: $v_1$ and $v_2$ as example, if the HTML content of the web page with pattern $v_1$ has one or more hyperlink (typically via HTML \texttt{<a>} tag) pointing to a URL of pattern $v_2$, then we say $v_1$ has an edge to $v_2$.
\end{itemize}

We show an example of sitemap for a simple online shopping site \cite{shah2016shoppingcart} in Figure \ref{fig_sitemap_example}. This sitemap shows the basic functionality of the site, including registration, login, product view, cart, etc.

There are several ways to build the sitemap for a site, including active crawling, passive sniffing and self-providing.

\paragraph{Active crawling.} Active crawling requires to run a crawler to build the sitemap of the site. The crawling typically starts from the homepage and enters each hyperlink from the current page recursively. Each URL pattern is retrieved only once to reduce the number of pages need crawling. This is based on the assumption that web pages with same URL pattern have the same page structure and similar hyperlinks of the same URL patterns.

\paragraph{Passive sniffing.} In passive sniffing, the URLs of site's traffic are monitored, learned and then used to build the sitemap. This scheme is less intrusive than active crawling. However, the sitemap may be incomplete limited by the amount of sniffed traffic.

\paragraph{Self providing.} The site provides its own sitemap for bot detection. By this way we could gain the most precise sitemap, but requires non-trivial work from the site owner.





\subsection{Subgraph Mapping}
\label{sec:subgraph_mapping}

\begin{figure}[t!]
	\centering
	\begin{subfigure}[b]{0.24\textwidth}
		\centering
		\includegraphics[width=1.7in]{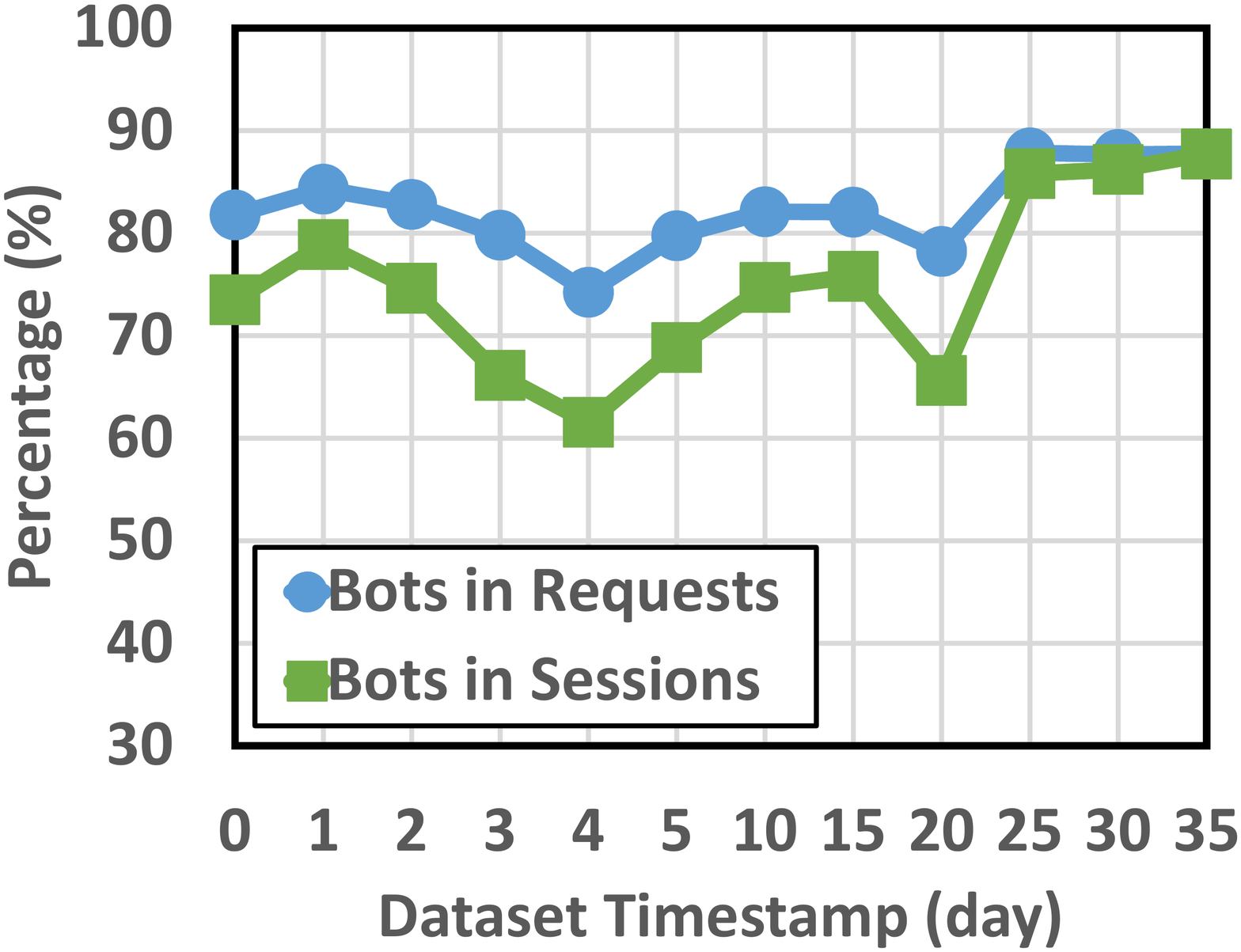}
		\caption{Number of bots.}
	\end{subfigure}%
	\begin{subfigure}[b]{0.24\textwidth}
		\centering
		\includegraphics[width=1.7in]{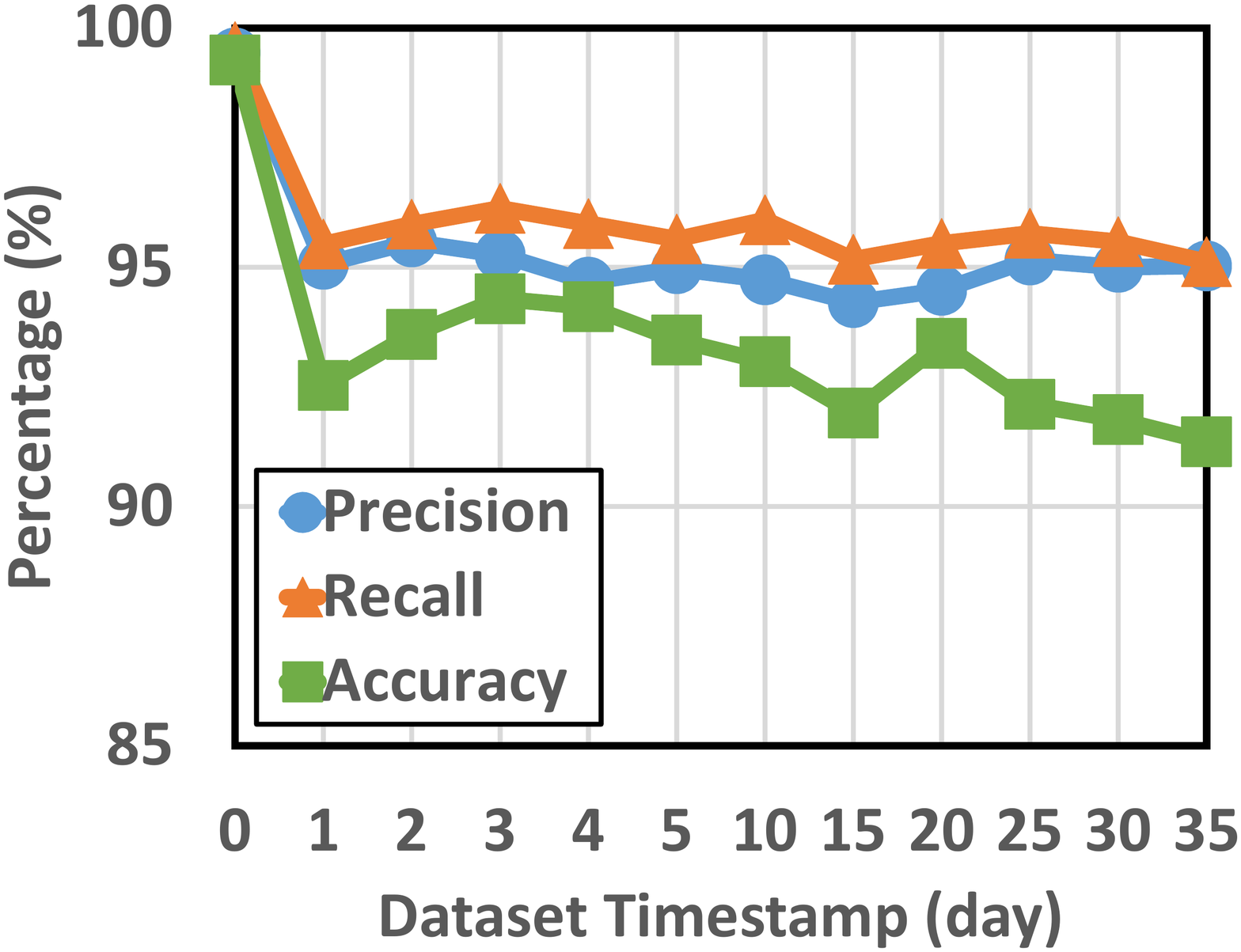}
		\caption{Precision, recall and accuracy.}
	\end{subfigure}
	\caption{Performance on search engine dataset spanning 35 days.}
\end{figure}

\begin{table}
	\centering
	\begin{tabular}{lrrl}
		\toprule
		Site & \#Nodes & \#Edges & How Is It Retrieved \\
		\midrule
		Search engine & 542 & 73441 & Self providing \\
		News site & 51 & 1473 & Active crawling \\
		University site & 134 & 2498 & \\
		\bottomrule
	\end{tabular}
	\caption{Sitemaps of our datasets.}
	\label{table_sitemap}
\end{table}

As shown in Figure \ref{fig_approach}, we can map \textit{RequestUri} of each request in a session into a node in the sitemap (in blue color). For each node in sitemap, we use the term ``access frequency'' to indicate the number of requests mapped to it. Moreover, two adjacent requests in the session can determine an edge in the sitemap (in solid line). Those mapped nodes and determined edges form a subgraph of the original sitemap. The generated subgraph contains information about the URL patterns the client has visited as well as the corresponding access frequencies.


There are occasions in which non-existent URLs are accessed and status code \textit{404} is responded (see Request 6 in Figure \ref{fig_approach}). This could be blamed for the bot's attempts to access a previously crawled and cached URL which has already been removed, or just brute-force attacks against vulnerable URLs like \textit{/backup.zip} and \textit{/apmserv5.2.6.rar}. These URLs cannot be mapped to any node in the sitemap. To resolve this, a node called ``INVALID'' is manually added to the sitemap as a container of all non-existent URLs.

\subsection{Image Generation}

In this section, we discuss about how to generate the trace image from a given subgraph in the section above. There are two kinds of elements in the trace image: spot and line. The procedure of image generation is described as follows: First we draw the black-filled spot for each node in the subgraph, the central coordinate and radius of which will be dwelt on later. Then we draw the straight line to bridge each pair of nodes containing an edge in original subgraph, the width of which is a session-free constant.

\subsubsection{Page Affinity}
\label{sec:coordinate}

To determine the position of each node in the sitemap, we use the Verlet algorithm \cite{verlet1967computer} to generate their coordinates. The Verlet algorithm performs molecular dynamics simulation based on Newton's equation of motion. The intuition is that a link between two nodes in the sitemap generates an attractive force, which tends to make them nodes closer. The algorithm use the iteration formula to reach the final balanced states for all nodes. In this way, the affinity of original web pages is visually exhibited on the generated sitemap.

It is notable that the coordinates are generated only once for a sitemap (site) and shared by all the subgraphs derived from this sitemap, which would not cause significant overhead when processing lots of sessions.

\subsubsection{Access Frequency}
\label{sec:access_frequency}

\begin{table}
	\centering
	\begin{tabular}{cllll}
		\toprule
		Client & \multicolumn{3}{c}{Trace Images} \\
		\midrule
		&
		\multirow{3}{*}{\includegraphics[width=0.5in]{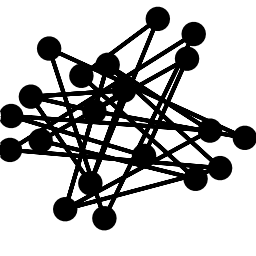}} &
		\multirow{3}{*}{\includegraphics[width=0.5in]{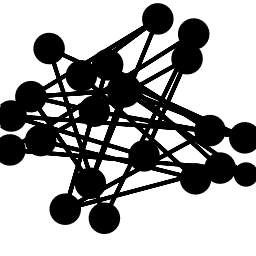}} &
		\multirow{3}{*}{\includegraphics[width=0.5in]{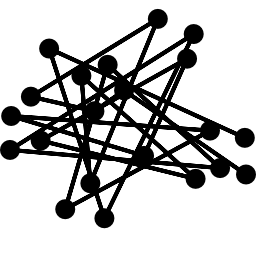}} &
		\multirow{3}{*}{\includegraphics[width=0.5in]{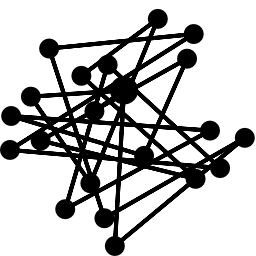}} \\
		Media- \\
		metribot \\
		
		&
		\multirow{3}{*}{\includegraphics[width=0.5in]{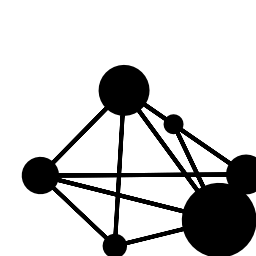}} &
		\multirow{3}{*}{\includegraphics[width=0.5in]{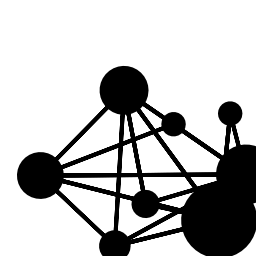}} &
		\multirow{3}{*}{\includegraphics[width=0.5in]{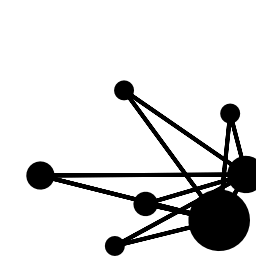}} &
		\multirow{3}{*}{\includegraphics[width=0.5in]{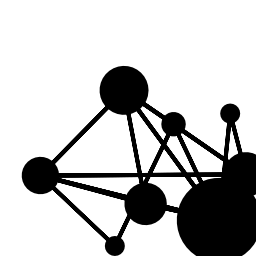}} \\
		\\
		AhrefsBot \\
		\\
		
		&
		\multirow{3}{*}{\includegraphics[width=0.5in]{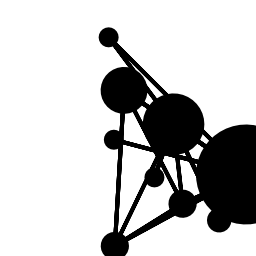}} &
		\multirow{3}{*}{\includegraphics[width=0.5in]{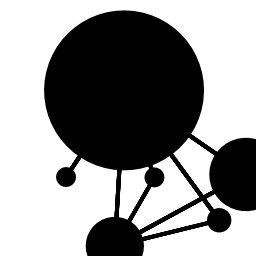}} &
		\multirow{3}{*}{\includegraphics[width=0.5in]{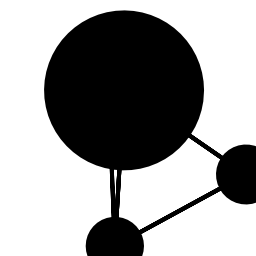}} &
		\multirow{3}{*}{\includegraphics[width=0.5in]{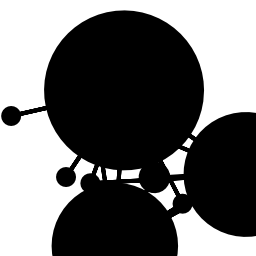}} \\
		\\
		Bingbot \\
		\\
		
		&
		\multirow{3}{*}{\includegraphics[width=0.5in]{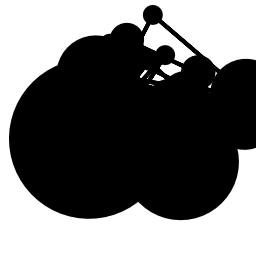}} &
		\multirow{3}{*}{\includegraphics[width=0.5in]{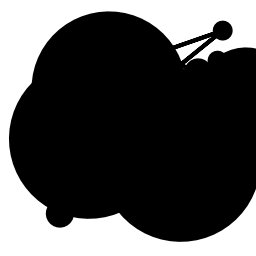}} &
		\multirow{3}{*}{\includegraphics[width=0.5in]{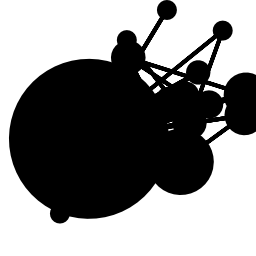}} &
		\multirow{3}{*}{\includegraphics[width=0.5in]{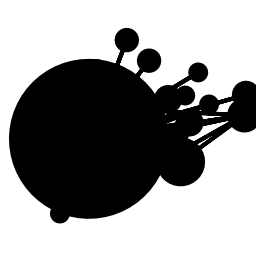}} \\
		\\
		Googlebot\\
		\\
		
		\midrule
		
		&
		\multirow{3}{*}{\includegraphics[width=0.5in]{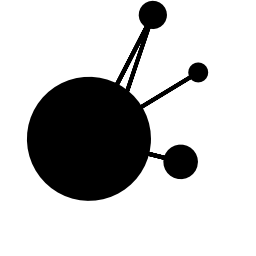}} &
		\multirow{3}{*}{\includegraphics[width=0.5in]{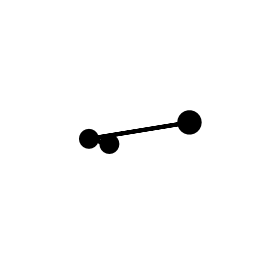}} &
		\multirow{3}{*}{\includegraphics[width=0.5in]{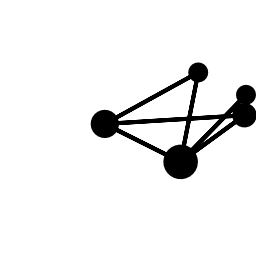}} &
		\multirow{3}{*}{\includegraphics[width=0.5in]{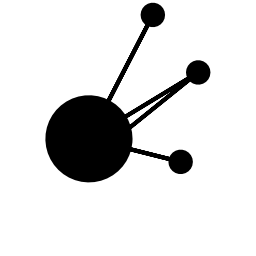}} \\
		\\
		\\
		Normal \\
		
		Users &
		\multirow{3}{*}{\includegraphics[width=0.5in]{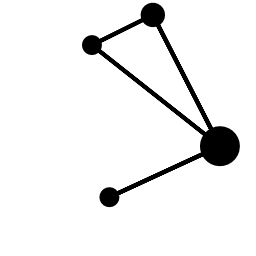}} &
		\multirow{3}{*}{\includegraphics[width=0.5in]{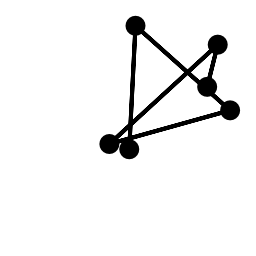}} &
		\multirow{3}{*}{\includegraphics[width=0.5in]{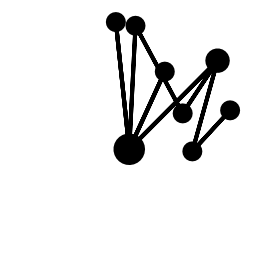}} &
		\multirow{3}{*}{\includegraphics[width=0.5in]{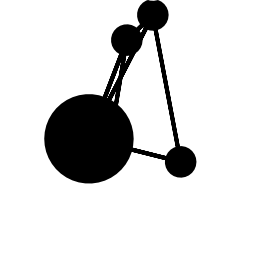}} \\
		\\
		\\
		\\
		\bottomrule
	\end{tabular}
	\caption{Trace images of different sessions.}
	\label{table_bot_image}
\end{table}

\begin{table*}
	\centering
	\begin{tabular}{llrrrrrrr}
		\toprule
		\multirow{2}{*}{Dataset} & \multirow{2}{*}{Type} & \multicolumn{2}{c}{Requests} & \multicolumn{2}{c}{Sessions} & Precision & Recall & Accuracy \\
		\cline{3-6}
		\rule{0pt}{2.2ex} & & \#Total & \textit{BoR} (\%) & \#Total & \textit{BoS} (\%) & (\%) & (\%) & (\%) \\
		\midrule
		Search engine (d: 0, hr: 0) & Train & 8212838 & 81.8 & 163811 & 73.5 & 99.5 & 99.6 & 99.3 \\
		Search engine (d: 5, hr: 0) & Test & 7541850 & 79.8 & 153464 & 68.8 & 95.0 & 95.6 & 93.5 \\
		News site (d: 0-3) & Train & 48606 & 78.4 & 4723 & 49.2 & 98.4 & 98.4 & 97.4 \\
		News site (d: 4-7) & Test & 52343 & 79.2 & 5070 & 51.6 & 98.0 & 96.9 & 95.7 \\
		University page (hr: 0-5) & Train & 250000 & 6.7 & 3837 & 7.5 & 97.8 & 93.8 & 99.4 \\
		University page (hr: 6-15) & Test & 250000 & 1.4 & 3754 & 4.7 & 95.3 & 69.1 & 98.4 \\
		\bottomrule
	\end{tabular}
	\caption{Performance on different datasets. \textit{BoR} indicates the bot's percentage of the total requests. \textit{BoS} indicates the bot's percentage of the total sessions.}
	\label{table_result}
\end{table*}

The access frequency for each URL pattern is an important feature for bot detection as a bot usually needs to access certain type of pages repeatedly, e.g., scraping product description pages like \textit{/product?id=*}. We use the spot radius to represent such access frequency for a sitemap node. Define $r=f(x)$, $r$ is the radius of the spot, $x$ is the access frequency. $f(x)$ is supposed to meet the following rules:

\begin{itemize}
	\item Higher access frequency results in larger node. So $f(x)$ is an increasing function.
	\item The smallest node (accessed only once) should also be visible in the image. Thus we have $f(1) = r_{min}$.
	\item The largest node should not occupy too much space of the image. Thus we have $f(+\infty) = r_{max}$.
	\item The gradient of $f(x)$ should be gentle when $x$ is relatively small. We can use $f(x_{gate}) = r_{gate}$ to restrict it. The $gate$ is a chosen value.
\end{itemize}




Inspired by the sigmoid function, we design our own $f(x)$ as follows. Given $r_{min}$, $r_{max}$, $x_{gate}$, $r_{gate}$, the parameters of $a$, $b$, $c$ can be determined by solving the constrains above.

\begin{align}
f(x) = \frac{c}{1 + e^{b-ax}}
\end{align}

\section{Evaluation}
\label{sec:evaluation}

\begin{table}
	\centering
	\begin{tabular}{llrr}
		\toprule
		Device & Type & TH (session/s) & LA (ms) \\
		\midrule
		Intel Xeon E5620 & CPU & 3.67 & 103.60 \\
		Intel i7-8750H & & 8.76 & 50.44 \\
		Intel i7-7700 & & 9.06 & 48.81 \\
		GTX 1050 Ti & GPU & 132.25 & 5.97 \\
		Tesla K40 & & 220.84 & 1.62 \\
		\bottomrule
	\end{tabular}
	\caption{Efficiency on the news site dataset under different CPUs and GPUs. \textit{TH} is training throughput. \textit{LA} is inference latency.}
	\label{table_cpu_gpu}
\end{table}

\begin{table}
	\centering
	\begin{tabular}{llrr}
		\toprule
		Scheme & Precision (\%) & Recall (\%) & Accuracy (\%) \\
		\midrule
		SVM & 82.4 & 32.6 & 97.9 \\
		XGBoost & 71.0 & 92.5 & 68.9 \\
		AdaBoost & 71.0 & 92.5 & 68.9 \\
		DT & 68.8 & 100.0 & 68.8 \\
		RF & 68.8 & 100.0 & 68.8 \\
		MLP & 73.8 & 87.8 & 70.2 \\
		LSTM+ & 35.0 & 81.4 & 95.1 \\
		SVM+ & 75.8 & 94.5 & 75.5 \\
		XGBoost+ & 86.8 & 98.9 & 88.8 \\
		AdaBoost+ & 85.4 & 96.7 & 86.3 \\
		DT+ & 85.9 & 98.7 & 87.9 \\
		RF+ & 83.8 & 96.5 & 84.8 \\
		MLP+ & 84.0 & 88.6 & 80.6 \\
		BotGraph & 95.1 & 95.5 & 92.5 \\
		\bottomrule
	\end{tabular}
	\caption{Comparison with other bot detection methods on the search engine dataset. The trailing \textit{+} indicates the \textit{clientIp} field is used as a feature.}
	\label{table_compare}
\end{table}

\subsection{Dataset}

To evaluate BotGraph, we use datasets from real world web server logs including Bing search engine and several sites from different industries like news site, university homepage, etc. As shown in the \texttt{Dataset} and \texttt{Type} columns, the training set of the search engine dataset is collected for a hour on day 0. The testing set is for the same hour on day 5. These logs have already been sessionized by tracking the \textit{SessionId} cookie of the client. Each session is labeled as \textit{bot} or \textit{non-bot}. The dataset only includes sessions the images of which have more than 3 spots, which will be explained in Section \ref{sec:discussion}. The labeling is performed as such: a team of 30+ professional engineers manually analyzed the traffic and used various ways including JavaScript support checking, mouse movement and click tracking, IP reputation, UserAgent blacklisting to label the traffic. In this paper, the labels are assumed to be accurate and used as ground truth.

\subsection{Setup}

\begin{table*}
	\centering
	\begin{tabular}{llllllll}
		\toprule
		Category & \multicolumn{6}{c}{Trace Images} \\
		\midrule
		&
		\multirow{3}{*}{\includegraphics[width=0.5in]{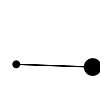}} &
		\multirow{3}{*}{\includegraphics[width=0.5in]{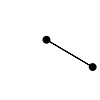}} &
		\multirow{3}{*}{\includegraphics[width=0.5in]{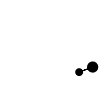}} &
		\multirow{3}{*}{\includegraphics[width=0.5in]{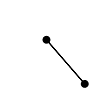}} &
		\multirow{3}{*}{\includegraphics[width=0.5in]{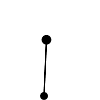}} &
		\multirow{3}{*}{\includegraphics[width=0.5in]{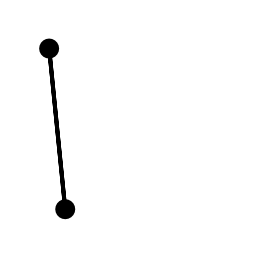}} &
		\multirow{3}{*}{\includegraphics[width=0.5in]{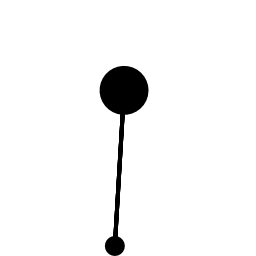}} \\
		False Positives & & & & & \\
		& & & & & \\
		
		&
		\multirow{3}{*}{\includegraphics[width=0.5in]{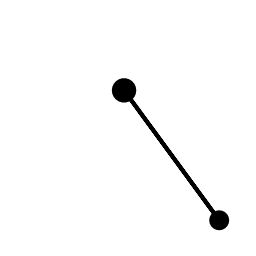}} &
		\multirow{3}{*}{\includegraphics[width=0.5in]{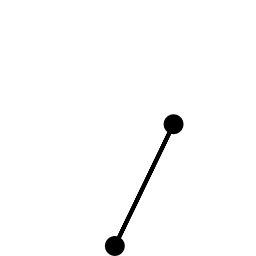}} &
		\multirow{3}{*}{\includegraphics[width=0.5in]{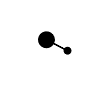}} &
		\multirow{3}{*}{\includegraphics[width=0.5in]{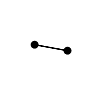}} &
		\multirow{3}{*}{\includegraphics[width=0.5in]{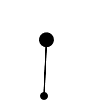}} &
		\multirow{3}{*}{\includegraphics[width=0.5in]{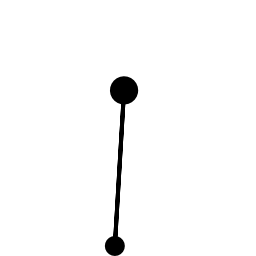}} &
		\multirow{3}{*}{\includegraphics[width=0.5in]{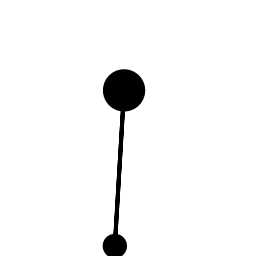}} \\
		& & & & & \\
		False Negatives & & & & & \\
		& & & & & \\
		\bottomrule
	\end{tabular}
	\caption{Trace images of false positives \& false negatives.}
	\label{table_fp_fn_image}
\end{table*}

The sitemaps for each dataset are shown in Table \ref{table_sitemap}. The search engine dataset provides a \textit{PageName} field, e.g., \textit{Home}, \textit{Page.Serp}, \textit{Page.NoResults}, \textit{Page.Image.Results}, etc. So we directly use it as the node in sitemap. Some random edges are generated to make most of the nodes connected in the sitemap. For other datasets, we used a crawler to actively scrape their sitemaps. It is notable that the sitemap edges have nothing to do with the subgraph edges. The sitemap edges are site-wise and only used as input of Verlet algorithm to generate the nodes' coordinates. However, subgraph edges are session-wise and used as a feature in the generated trace images.


We use $256 \times 256$ as the image size. A padding of 5\% is added to the image's four sides to ensure the spots cannot exceed the canvas easily. We empirically use $r_{min} = 4, r_{max} = 80, x_{gate} = 50, r_{gate} = 50$ in the access frequency function $f(x)$. The parameters of $f(x)$ can be solved accordingly.



To classify the trace images, we tried different CNN models, including LeNet-5, AlexNet, ResNet, etc. They all get similar precision and recall. Thus we choose the fastest LeNet-5, a 7-level CNN to train and inference. The trace images are used as input. The output is a scalar, indicating bot or non-bot. We use the following hyperparameters in our experiments: $Batch Size = 64, Epoch = 100, LR = 0.01, SGD Momentum = 0.5$. Our training code is open sourced at: \url{https://github.com/botrainer/botrainer}.

%



\subsection{Performance}

Although a bot can easily modify its \textit{UserAgent} to pretend a normal browser, A client with \textit{UserAgent} claiming to be a bot is usually for real. So we use the claimed \textit{UserAgent} of each trace image (aka session) as a group key and show several groups of randomly selected trace images in Table \ref{table_bot_image}. We find these images have described the client's behavior pretty well in the following two aspects:


\begin{enumerate}
	\item The trace images of the same bot share high similarity. Different bots have distinct image patterns.
	\item The trace images of normal users have different shapes, which are usually not the same as those of bots.
\end{enumerate}




The performance of BotGraph on different datasets is shown in Table \ref{table_result}. besides precision, recall and accuracy, we also present two interesting metrics which are related to the bot detection result: bot's percentage of requests ($BoR$) and bot's percentage of sessions ($BoS$). $BoR$ and $BoS$ are usually not the same value but highly related. When $BoS \geq 49\%$, we usually have $BoR > BoS$. This is because the session length (number of requests) of a bot is larger than that the session length of a normal user on average. So when bots are significant in a traffic, this pattern is more obvious. However, when $BoS < 10\%$, the existing bots are usually unorganized and with no harmful intention, like Googlebot, Bingbot, etc. They do not crawl very large number of pages in a session, which is not significantly different from normal users. For the datasets with $BoS \geq 49\%$, BotGraph achieves higher than 95\% precision and recall. It means when a site is heavily affected by bots, BotGraph can effectively detect those bot traffic. For the datasets with $BoS < 10\%$, BotGraph still gets higher than 95\% precision with nearly 70\% recall. We think it is because the bot traffic in such sites are scattered and have no stable patterns, which influences our effect.


Our CNN-based model for image classification is implemented in PyTorch. We benchmarked the training and inference performance in different circumstances on the previously mentioned news site's dataset, as shown in Table \ref{table_cpu_gpu}. We can see that under an ordinary GPU, BotGraph does not cause obvious latency compared to the common round-trip delay of 50\url{~}100 ms on Internet.


\subsection{Comparison}

We compared BotGraph with other bot detection methods like long short-term memory (LSTM), SVM, XGBoost \cite{ldaevaluation}, AdaBoost, decision tree (DT), random forest (RF), multi-layer perceptron (MLP), etc. Some methods are proposed in previous work and some others are implemented by ourselves. The experiment is done on the dataset of news site, as shown in Table \ref{table_compare}. we also present the request fields that each method uses. We use the following feature engineering approach for each field: \textit{UserAgent} is parsed into tuple $(Browser, OS, Device)$. \textit{ClientIp} (IPv4) is directly used in its 32-bit integer form. For fairness, we also add the session length as a feature, which is similar to the access frequency of BotGraph. We provide both results with and without the \textit{ClientIp} feature. It shows that most methods have no more than 75\% precision without \textit{ClientIp} and 87\% precision with \textit{ClientIp}. It is also notable that BotGraph uses neither \textit{ClientIp} nor \textit{UserAgent} as features, but achieves $\sim$95\% precision and recall.

%

\section{Discussion}
\label{sec:discussion}

A weakness of BotGraph is when the number of spots in trace images is pretty small (e.g., $<$ 3), the detection result can be wrong. It is because the normal users and bots are more likely to have similar page browsing behaviors when only one or two page patterns are accessed. We show several randomly selected false positives and false negatives in Table \ref{table_fp_fn_image}. We can see some images of false positives are nearly the same as the images of false negatives. We believe this drawback is not severe, considering bots are harmful largely due to their large amount of traffic caused to the site. In fact, we found that for more than 95\% of the sessions with 1$\sim$2 spots, their access frequency is also minimized by accessing each page only once. We think a bot which only makes one or two requests in total will do no harm to the site.


\section{Conclusion and Future Work}
\label{sec:conclusion}

Website bots have been proven to be a severe threat for Internet these years. In this paper, BotGraph provides a novel scheme to describe bot behaviors in 2-dimensional images. Then state-of-the-art image classification methods like CNN can be used to determine whether a sesison is a bot. The experiments on real-world 35-day datasets show that BotGraph is a very effective model to detect bots: it achieves $\sim$95\% both in precision and recall. BotGraph leverages the client's behavior instead of its identity as features, it is a promising way to detect advanced bots that frequently change their identities. Currently, we are working on more generic graph-based classification method to take advantage of underlying general graph related feature to better describe the characteristics of bots.


%

\appendix

\bibliographystyle{named}
\bibliography{cnn}

\end{document}